# Study of Electron Transport in Organic and Inorganic Atomic Monolayer Based MOS/MOSFET

J. Cyril Robinson Azariah, U. Satheesh, D. Devaprakasam[*]

*NEMS/MEMS and Nanolithography Lab, Department of Nanosciences and Technology, Karunya University, Coimbatore 641114. Email: devaprakasam@karunya.edu*

**Abstract:**

The wide research interest for the potential nanoelectronics applications are attracted by the organic and inorganic monolayer materials. In this work, we have studied the organic monolayer such as trichloro (1H,1H,2H,2H-perfluorooctyl)-silane (FOTS), hexamethyldisilazane (HMDS) and inorganic monolayers such as hexagonal - boron nitride (h-BN) and molybdenum disulfide (MoS2) based MOS devices. The organic monolayer based configurations are Au/FOTS/p-Si and Au/HMDS/p-Si. The inorganic monolayer based configurations are Au/MoS2/SiO2/p-Si and Au/h-BN/SiO2/p-Si. These configurations were examined and compared with Au/SiO2/p-Si MOS configuration using the Multi-dielectric Energy Band Diagram Program (MEBDP) and MOSFeT simulation software. The C-V and I-V characteristics of MOS and MOSFET of FOTS, HMDS, h-BN, MoS2 and SiO2 were reported. The results show that the above configurations are suitable for designing MOSFETs with smaller drain induced barrier lowering (DIBL) and reduced threshold voltage. We noted that the above configurations are better than 2nm thick dielectric SiO2 based MOSFET with a channel length of 10nm.

## 1. Introduction

In the past fifty decades of the silicon industry, the thickness of SiO2 has been aggressively scaled towards 1.2nm (90nm node technology) and 0.8nm (45nm technology). Further reduction in the thickness of SiO2 leads to leakage current in the device and second order effects like short channel effect, tunneling of currents, impact ionization and so on. In order to overcome the above disadvantages, we go for replacement of SiO2. We understood that organic/inorganic based monolayer molecules can be utilized in the thin-film transistor (TFT) gate dielectric research [1]. The applications using these monolayer molecules includes: (i) very low gate leakage currents and very good chemical/thermal stability, (ii) reduced interface trap state densities to maximize charge transport efficiency, (iii) compatibility with both p-channel and n-channel organic/inorganic semiconductors, (iv) enhanced Capacitance vs. Voltages with dynamic characteristics, and (v) efficient fabrication via solution-phase processing methods.

In this work, we focus on the substitute gate dielectric materials monolayers of (i) organic molecules such as organosilanes like FOTS [1-3] and HMDS [1] and (ii) inorganic molecules such as 2D dichalcogenide like hexagonal-BN [4] and MoS2 [5-16] which are found to have very good insulating properties, large capacitance values and essential properties for addressing these challenges.

The outline of the paper is as follows. In the second section, we define the simulation approach in detail. In the third section, we present the results of simulation and the discussions. At last, we briefly summarize the results and conclusions of this paper.

## 2. Approach

The simulated device structure [20] is illustrated in Figure 1. It is a structure representing the Metal-Oxide-Semiconductor Field Effect Transistor (MOSFET). FOTS, HMDS, h-BN and MoS2 are used for the gate insulator with a thickness of *tox* = 2nm. These monolayers are used as the channel materials. The source (drain) is heavily doped with a doping concentration of $8.5 \times 10^{17} m^{-2}$, and its length is *LS* (*LD*) = 10nm8. The channel is intrinsic, and its length is equal to the length of gate *LG*. n-type MOSFET is fabricated on the p-type silicon substrate. The source and drain regions are formed by diffusing n-type impurities to form the depletion region.

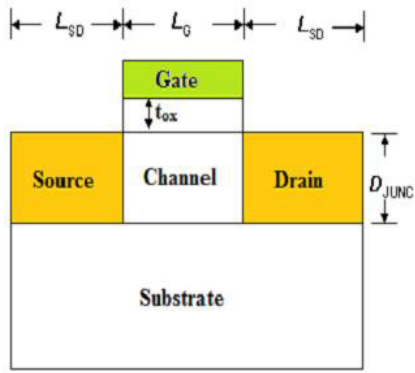

Figure 1 shows the schematic of the MOSFET device structure.

Table 1 shows the different materials energy gap and dielectric constant values. FOTS/HMDS/h-BN/MoS2 were used for the gate insulator with a thickness of *tox* = 2nm. The source (drain) is heavily doped with a doping density of $2\times10^{20}/m^2$, and its length is *LS* (*LD*) = 50nm in simulation. The length of the intrinsic channel is equal to the length of gate *LG* which varies from 3nm to 20nm. Djunc is the junction Depth at the channel/drift region.

|         | Si   | MoS2 | HMDS | FOTS | h-BN | SiO2 |
|---------|------|------|------|------|------|------|
| Eg (eV) | 1.12 | 1.9  | 2.45 | 5    | 5.2  | 9    |
| κ       | 11.8 | 6    | 2.2  | 4    | 5    | 3.9  |

Table I Energy-gap (Eg) and dielectric constant (κ) of various materials

## Results and Discussions

Using the MEBDP simulation, the oxide capacitances are calculated for the various dielectrics used in the MOS.

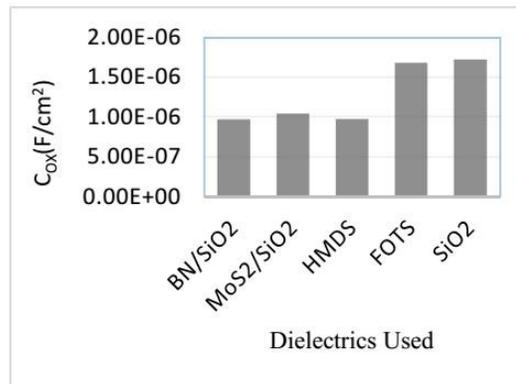

Figure 2 shows the oxide capacitance of various MOS structures

The C-V characteristics of the organic and inorganic monolayer based FOTS/HMDS/h-BN/MoS2 MOS structures is shown in Figure 3 (a) and 3(b).

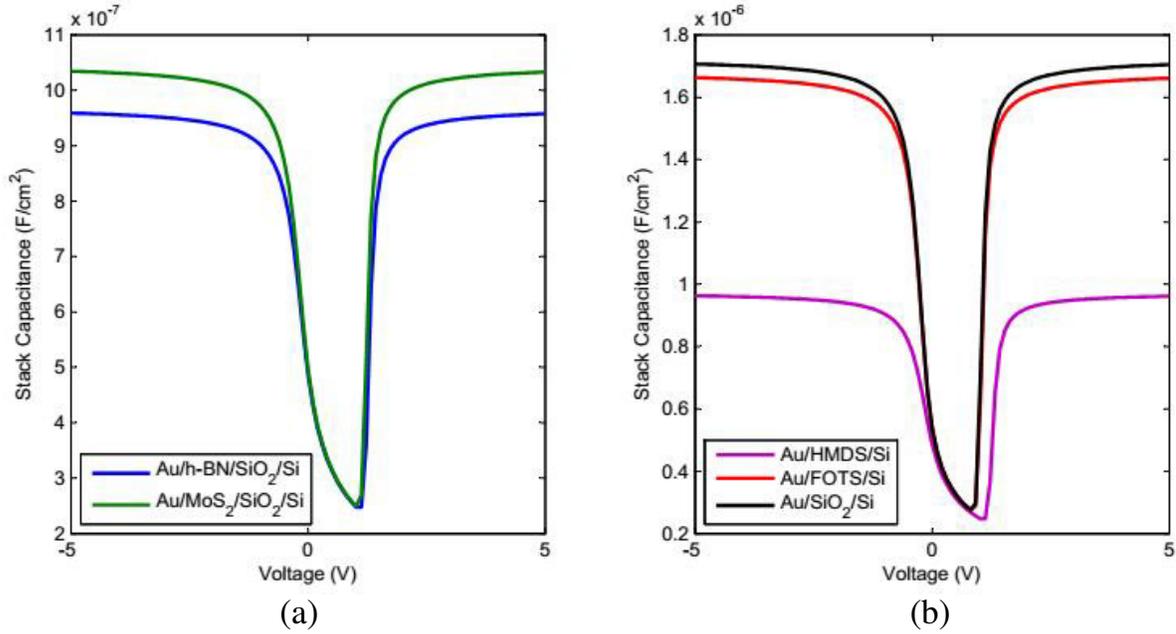

(a)     (b)

Figure 3 (a) and (b) shows the comparison of the C-*V* characteristic of the different monolayer MOS device configurations. The gate length *LG* is set to 10nm. Au used as gate material and its work function is used to achieve the specific off-current (100nA/$\mu$m).

From the characteristics, we observed that the inorganic monolayers exhibits the maximum capacitance and the organic monolayers exhibits the minimum capacitance. The dynamic Qg-V characteristics of the device are studied using the simulation plot between the gate-charge versus the applied voltage characterization.

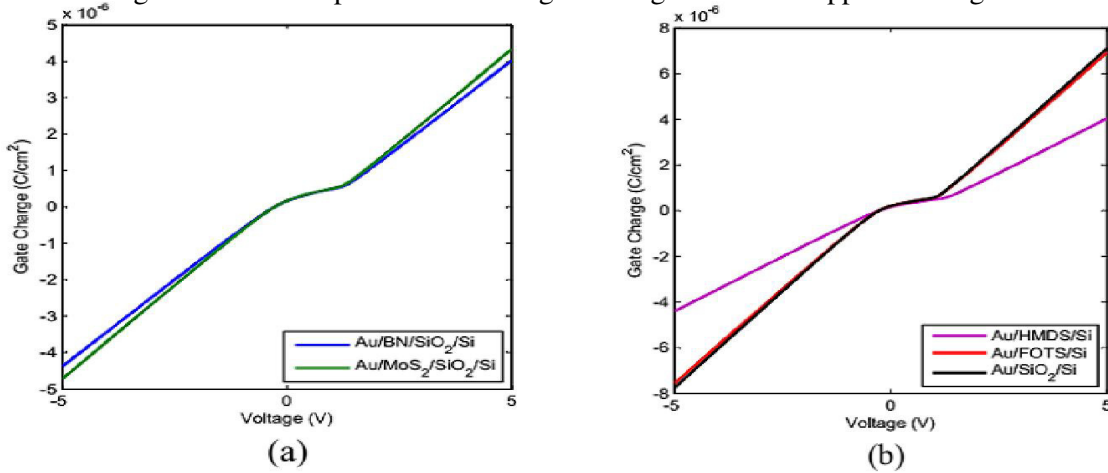

(a)     (b)

Figure 4 (a) & (b) shows the characteristics curve drawn between gate charges versus applied voltage of the various monolayer MOS structures.

Figure 4(a) shows the characteristics curve drawn between gate charge and the applied voltage between -5 V to +5 V for the various inorganic monolayer MOS structures. Figure 4(b) shows the characteristics curve drawn between gate charge and the applied voltage between -5 V to +5 V for the various organic monolayer MOS structures compared with the SiO2 based MOS structure. In the MOS device, a voltage applied to the gate controls the current flow between the drain and the source. This gives the dynamic characteristics of the MOS device, which is clearly studied. The charging and the discharging of the MOS device are the dynamic behavior. It is simulated using the MEBDP simulator. [18][19]

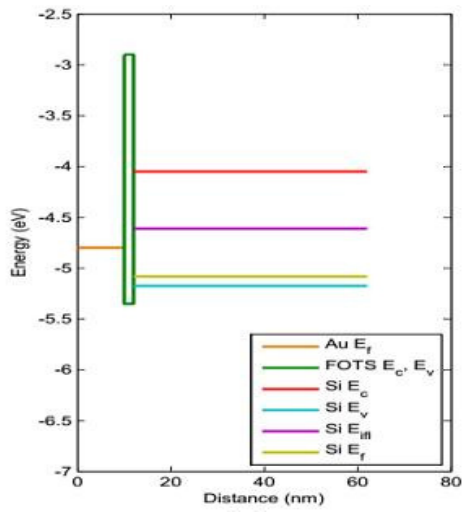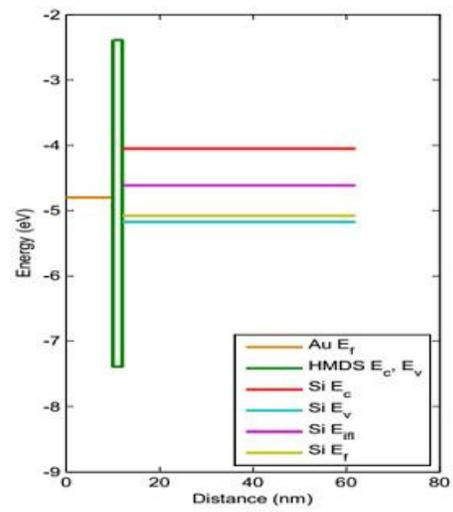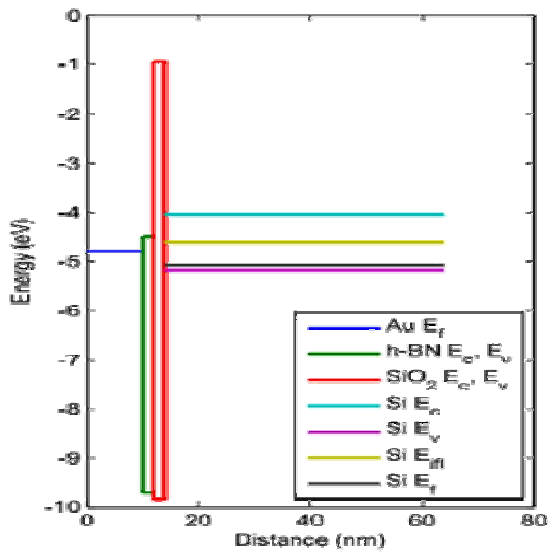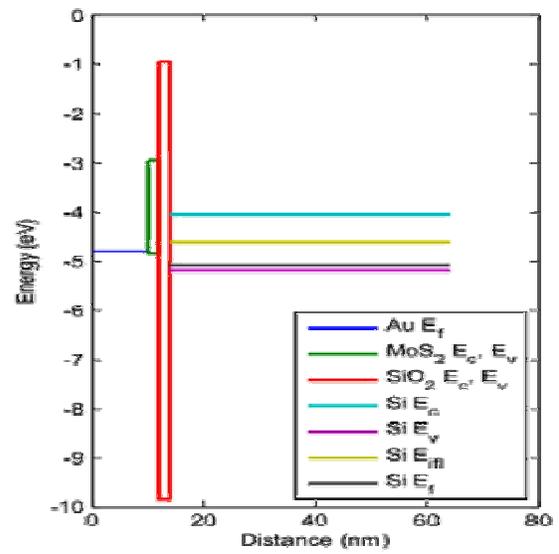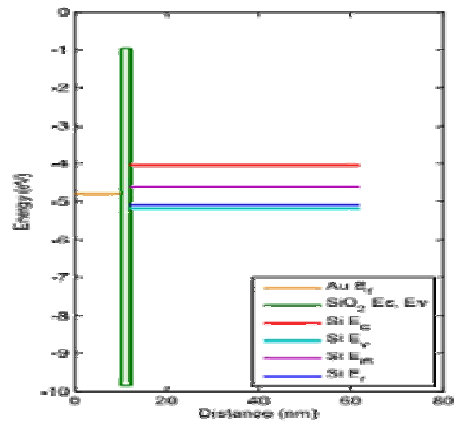

Figure 5 shows the energy band diagram of the monolayer MOS structures at the flat-band condition with $L_G$ = 10nm (a) FOTS, (b) HMDS, (c) h-BN, (d) MoS2 and compared with (e) SiO2.

Figure 5 illustrates the energy band diagram of the monolayer MOS structures with LG=10nm. From the energy band simulation at the flat-band condition, the oxide barrier height was calculated and is shown in Table 2.

Table 2 shows the oxide barrier height calculated for the various MOS structures involving monolayer dielectrics.

| Monolayer used | Oxide barrier height (eV) |
|---|---|
| FOTS | 2.41 |
| HMDS | 1.9 |
| h-BN | 0.3 |
| $MoS_2$ | 2.3 |
| $SiO_2$ | 3.85 |

We used the following parameters for the MOSFeT simulations: the silicon band gap of 1.12eV at 300K, its dielectric constant 11.8, electron saturation velocity $1.03*10^{+7}$cm/s, beta value set to 2 and the electron mobility is kept at 1400cm$^2$/Vs. The current conduction in a MOSFET is determined by the flow of electron from the source to the drain through the channel. When the gate source voltage is below the threshold voltage, the device remains in the cut-off region and does not conduct any current. With the smallest drain bias of Vd=0.05V, the current conduction initiated between the source and drain.

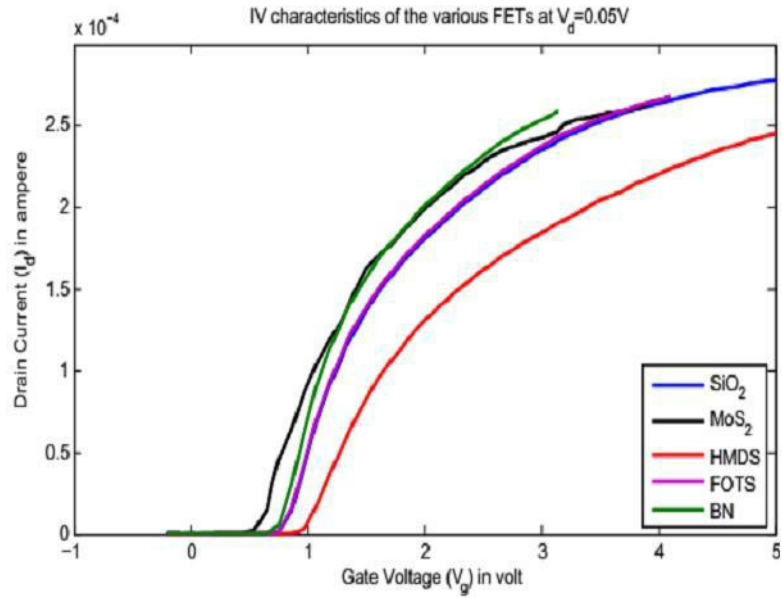

Figure 7 shows the combined IV characteristics of MOSFET at Vd=0.05V, whose gate dielectrics are SiO2 compared with the following monolayer based gate dielectrics MoS2, h-BN, FOTS and HMDS.

The combined IV characteristics of MOSFETS of different gate dielectric materials simulated at

Vd=0.05V, is shown in figure 7. It is observed that the threshold voltage for the inorganic monolayer h-BN and MoS2 are lower than SiO2 and the threshold voltage for the organic monolayer FOTS and HMDS is higher than the SiO2. When Vg>Vth, the drain current increases.

## 4. Conclusion

Hence, the CV characteristics of various MOS devices with the organic and inorganic monolayer dielectric membrane studied using the Multi-dielectric Energy Band Diagram Program. We also studied the IV characteristic of MOSFET configuration of the organic and inorganic membrane using the MOSFeT simulator. The results of organic and inorganic membrane based MOS and MOSFET devices MEBDP and MOSFeT simulations show that the above configurations are suitable for designing FETs with smaller Drain Induced Barrier Lowering (DIBL) and reduced threshold voltage. We noted that the above configurations are better than 2nm-thick-dielectric SiO2 FETs at a channel length of 10nm with the same gating.

### Acknowledgement

We thank DST - Nano mission, Government of India and Karunya University for providing the financial support to carry out the research. We also thank the department of Nanosciences and technology for the help and support to this research.